# Complementary LEEM and eV-TEM for imaging and spectroscopy

Peter S. Neu*[1], Daniël Geelen*[1], Aniket Thete[1], Rudolf M. Tromp[1,2] & Sense Jan van der Molen[1]

[1]Leiden Institute of Physics, Niels Bohrweg 2, Leiden, The Netherlands

[2]IBM T.J. Watson Research Center, Yorktown Heights, New York 10598, USA

*These authors contributed equally

## Abstract

Transmission electron microscopy at very low energy is a promising way to avoid damaging delicate biological samples with the incident electrons, a known problem in conventional transmission electron microscopy. For imaging in the 0-30 eV range, we added a second electron source to a low energy electron microscopy (LEEM) setup, enabling imaging in both transmission and reflection mode at nanometer (nm) resolution. The latter is experimentally demonstrated for free-standing graphene. Exemplary eV-TEM micrographs of gold nanoparticles suspended on graphene and of DNA origami rectangles on graphene oxide further establish the capabilities of the technique. The long and short axes of the DNA origami rectangles are discernable even after an hour of illumination with low energy electrons. In combination with recent developments in 2D membranes, allowing for versatile sample preparation, eV-TEM is paving the way to damage-free imaging of biological samples at nm resolution.

## 1. Introduction

Electron microscopy, in particular Transmission Electron Microscopy (TEM), has become a major tool for disciplines ranging from archaeology[1] to materials science[2] and biology[3]. Modern TEM instruments accelerate electrons to hundreds of kilo-electron volts[4], where the electron Mean Free Path (MFP) increases with energy, enabling atomic resolution[5,6]. However, at such high energies beam damage is a problem, especially in organic molecules and biological specimens. Proteins, for example, can already be altered by a single electron impact[7], so imaging at lower electron energies would be preferred. Recently, the Sub-Angstrom Low-Voltage Electron (SALVE) microscopy project[8] showed, that operating at 20 keV instead of 80 keV allows for a two orders of magnitude increase in dose when imaging fullerenes[9].

Strikingly, towards very low energies (i.e. below ~30 eV) the MFP also increases, as suggested by the so-called 'universal' MFP vs. energy curve[10]. This allows for high electron transmission of sufficiently thin samples, with the potential of nanometer (nm) spatial resolution with minor damage. The reason for the increase of MFP at very low energies is that fewer interaction processes, e.g.

plasmonic or excitonic, can be excited. Recently, electron microscopy techniques in this very low energy regime have been successful, with a scanning TEM instrument imaging graphene[11] and a holography type instrument imaging various microorganisms[12,13]. Here, we demonstrate the imaging capabilities of eV-TEM[14–16], a new form of TEM that illuminates the sample with a collimated beam of 0-30 eV electrons, i.e. an energy four to six orders of magnitude lower than conventional TEM.

## 2. eV-TEM setup

To acquire TEM micrographs at low energies, we have modified an Aberration-Corrected Low-Energy Electron Microscope (AC-LEEM) by placing a miniature electron gun (0-100 eV)[14] behind the sample (see Fig. 1a). Electrons emitted from the TEM-gun illuminate the sample along the optical axis in a collimated beam (green). Transmitted electrons follow the same optical path (red) as electrons from the cold field emission LEEM gun (black) upon reflection. Thus, images of the same area can be obtained by reflected (LEEM) and transmitted (eV-TEM) electrons, by utilizing either electron source.

The theoretical resolution of our system depends on the angular spread α of electrons contributing to the image, which can be adjusted with the contrast aperture (Fig. 1a). As our setup (named ESCHER[17]) includes an aberration correcting mirror, spherical aberrations are corrected to third order and chromatic aberrations to second rank. In the paraxial limit the aberration corrected resolution (full width at half maximum) is given by [18]

$$\text{FWHM} = \left\{ \left(\frac{0.61\lambda}{\alpha}\right)^2 + (C_5\alpha^5)^2 + \right.$$
$$\left. \left(\frac{\Delta E}{E}\right)^2 (C_{C3}\alpha^3)^2 + \left(\frac{\Delta E}{E}\right)^4 (C_{CC}\alpha)^2 \right\}^{1/2},$$

with electron wavelength $\lambda$ at the column energy $E$=15 keV, energy spread $\Delta E$ and the aberration coefficients $C_5$, $C_{C3}$ and $C_{CC}$ reported in [19]. At low acceptance angles the Rayleigh criterion (Fig. 1b, dashed line) limits the resolution, at high acceptance angles the spherical and chromatic aberrations are limiting. The calculation for 3 eV electron energy, shown in Fig. 1b, is based on the aberration coefficients in [19] and an energy spread of 0.25 eV and 0.8 eV for LEEM and eV-TEM, respectively. The energy spread of 0.8 eV in eV-TEM is typical for the thermal barium-oxide disk emitter used (Kimball Physics, ES-015), while an energy spread of 0.25 eV is characteristic for the cold-field emission LEEM gun.

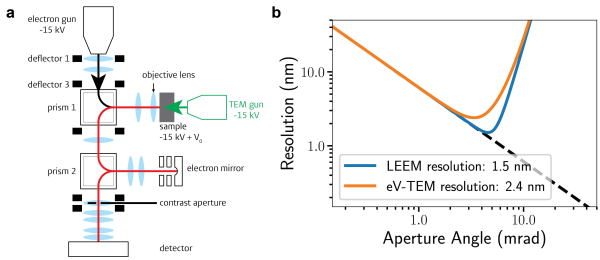

*Figure 1: Extension of an Aberration-Corrected Low-Energy Electron Microscope (AC-LEEM) with a second electron gun situated behind the sample to enable transmission electron microscopy at few-eV energies (eV-TEM). (a) Electron Microscopy setup for imaging a sample in both reflection mode (LEEM) and transmission mode (eV-TEM). In LEEM the electrons are directed towards the sample through the objective lens (black path); In eV-TEM the electron gun is situated on the other side of the sample (green path). The detection path (red) is the same in both modes. (b) Resolution in aberration corrected LEEM and eV-TEM depends on the acceptance angle of electrons, which is adjusted by the contrast aperture.*

## 3. Resolution on few-layer graphene

To determine the resolution of eV-TEM, a high-magnification micrograph of a few-layer graphene sample was recorded. For this, we first coated a holey silicon nitride (SiN) grid (2.5 µm holes, PELCO® Holey Silicon Nitride Support Film from Ted Pella) with 5 nm Pt/Pd. Using the polymer-free method described in [20], we then transferred graphene onto the metal-coated SiN grid. Inside the microscope vacuum chamber (pressure below $10^{-9}$ mbar), the sample was heated with a laser heater to outgas residues and contaminants.

In Fig. 2a we show an eV-TEM image of graphene, which exhibits stark contrast between areas of different thicknesses[16]. As graphene is highly conductive[21] and the work function is nearly constant[22,23] over the whole sample, the electric field between sample and objective lens (15 kV/1.5 mm) is perpendicular to the sample. This is desirable, as an in-plane electric fields (for instance due to large workfunction differences) would locally deflect the low energy electrons and deform the image.

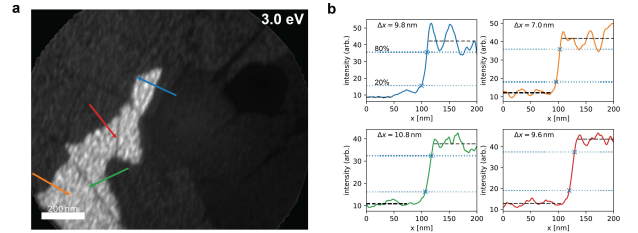

*Figure 2: Contrast between areas of graphene with different thicknesses caused by resonant electron transmission. (a) eV-TEM micrograph of a few-layer graphene sample at 3 eV electron energy. Intensity profiles (b) are extracted from the at the positions indicated in (a). The resolution $\Delta x$ is determined from the points where intensity rises from 20% to 80% of the spread between maximum and minimum intensity.*

From the image in Fig. 2a, we extract the intensity profiles (Fig. 2b), as indicated in the micrograph in the corresponding colors. Each line has a width of five pixels, over which the intensity is averaged perpendicular to the line. We define the resolution $\Delta x$ as the distance over which the intensity increases from 20% to 80% (dotted lines) of the intensity step from the dark to bright area. Along the different profiles we observe a resolution between 7 nm and 11 nm, showing that eV-TEM can reach sub-10 nm resolution. We believe that the resolution is limited by the relatively large energy spread of the source, and by the large emitting area of the cathode, limiting both temporal and spatial coherence. The current electron source is a 0.84 mm diameter barium-oxide covered refractory metal disc

(Kimball Physics), so replacing it by a laser-illuminated, low work-function cathode (such as $K_2CsSb$, cesiated graphene or GaAs) would improve the energy spread and reduce the emitting area.

The experimental values for $\Delta x$ are to be compared to the theoretical limit (Fig. 1b). The theoretical resolution limit of 2.4 nm (full width at half maximum) for 3 eV electrons at 0.8 eV energy spread (ignoring the spatial extent of the source) corresponds to $\Delta x = 1.7$ nm in terms of the 20%-80% criterion. The contrast aperture was chosen to optimize contrast in this experiment, rather than resolution, so it is likely to be too small for the resolution optimum (the minimum in Fig. 1b). Also, the alignment of the incoming electron beam cannot be adjusted during the experiment, as there is no deflector between eV-TEM electron gun and the sample. Additionally, in these experiments no systematic effort was made to exactly cancel the spherical and chromatic aberrations of the objective lens, so this undoubtedly also contributes to the observed resolution. In LEEM the microscope routinely shows sub-2 nm spatial resolution, so we anticipate that future improvements to the electron illumination system will result in a significant improvement of eV-TEM resolution.

## 4. Transmissivity and reflectivity of few-layer graphene

Spectra were measured by scanning the electron energy from 0 to 75 eV and recording an image every 0.1 eV in LEEM and eV-TEM mode. From these datasets, we extract reflectivity (Fig. 3a) and transmissivity (Fig. 3b) spectra of the mono-, bi- and trilayer graphene areas indicated in the inset figures.

The spectra exhibit characteristic maxima and minima, which can be explained to a large extent by a simple toy model inspired by thin film optics[24]. In that model, the splitting of the reflectivity minimum (transmissivity maximum) around 5 eV has been understood as the destructive (constructive) interference of the electron wave function as it is partially reflected and transmitted from each graphene layer[16]. Therefore, the monolayer spectra do not exhibit oscillations in this energy range, whereas each additional layer gives rise to one additional minimum. By scanning the energy, the electron wavevector k is varied and the reflectivity minima (transmissivity maxima) form when the interference condition $k \cdot 2d = (2n + 1) \cdot \pi$ (with layer spacing d and integer n) is met[16]. The reflectivity minimum (transmissivity maximum) just below 20 eV is attributed to the next order interference pattern. At this energy, the expected splitting of the reflectivity minimum for the 3-layer sample is not discernible due to inelastic broadening [16,25], although the 3-layer minimum is much broader than the 2-layer minimum as a result of this splitting. The next sets of reflectivity minima (at 30 eV and 50 eV) can no longer be explained by the simple toy model   but are accounted for by *ab initio* theory [26,27].

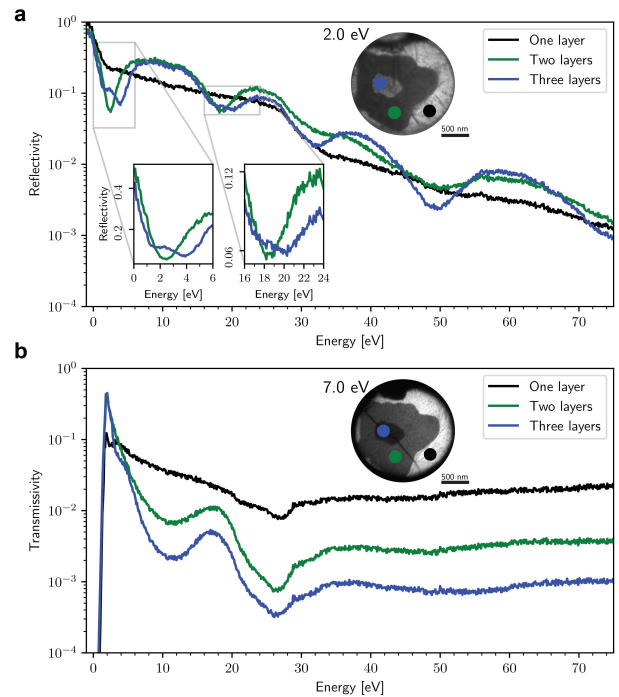

*Figure 3: Low energy electron reflection (a) and transmission (b) spectra for one, two and three layers of free-standing graphene. The circles in the inset images show the areas where the spectra were extracted. The splitting of the reflectivity minimum around 5 eV (inset in a) is characteristic of the layer count. The drop in both reflectivity and transmissivity at 28 eV is attributed to the formation of the diffracted beams, which are blocked by the contrast aperture.*

A drop in reflectivity and transmissivity at ~ 28 eV is visible for the monolayer as well as the bilayer and trilayer spectra. This reduction in both reflectivity and transmissivity is caused by the appearance of the first-order diffracted beams, which are blocked by the contrast aperture and do not contribute to the collected intensity. The intensity already drops before the diffracted beams are visible in LEED experiments (above 33 eV), because the diffracted electrons have to overcome the work function of 4.6 eV ($\pm 0.1$ eV in the different layer counts)[22,28] to reach the vacuum. Thus at 28 eV incident energy, the diffracted beams are formed but confined to the interior of the material, therefore lowering the intensity of the zeroth order spot.

If we consider the broader trend over the probed energy range, we find that the reflectivity decreases roughly exponentially with energy, whereas the transmissivity reaches a minimum at approx. 28 eV. The increase in transmissivity towards higher energies signifies a slowly increasing MFP. The overall trend in transmissivity (disregarding resonant oscillations) is in qualitative agreement with the 'universal' MFP curve, which estimates the minimal MFP to be at about 30-40 eV. However, we also stress that the layer-oscillations seen below 20 eV are in striking disagreement with this 'universal' curve[16].

## 5. Few-Layer materials as a high-transmissivity support film for imaging

Thin support films based on layered materials are common in TEM[29] due to their high electron-transparency and low scattering contrast. Here, we explore their use in eV-TEM. Support films in eV-TEM are required to be flat, therefore the layers have been prepared on SiN perforated TEM support grids as described above.

### 5.1 Gold nanoparticles on a graphene substrate

As a first high-contrast example, we imaged gold nanoparticles (10 nm diameter) in both LEEM (Fig. 4a) and eV-TEM (Fig. 4c). Note that imaging the same area with reflected and transmitted low energy electrons is conveniently possible by switching between the two electron sources. Samples were prepared by placing a ~ 16 µl droplet of Au NP in water solution on suspended graphene and leaving it to dry overnight. With this method it is well possible that a few NPs form a somewhat larger cluster. After transferring to the vacuum setup, the sample was first heated to 440°C for 90 minutes, and then cooled down to room temperature for imaging.

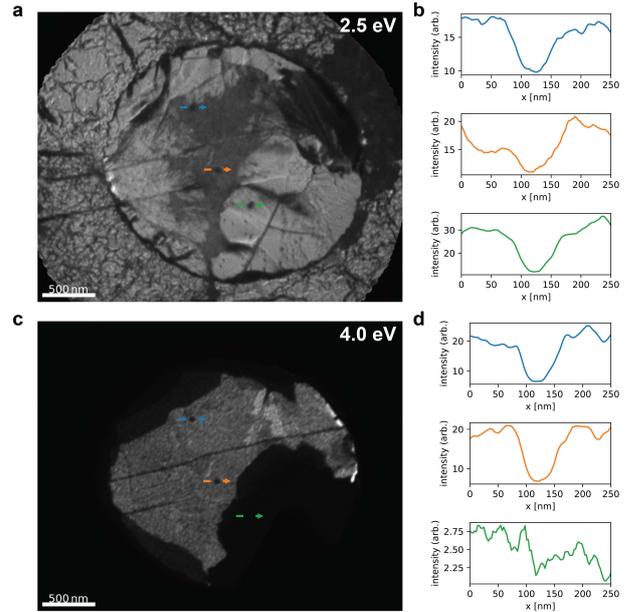

Figure 4: Gold nanoparticles on top of suspended graphene membrane. LEEM (a) and eV-TEM (c) image of same area with Au nanoparticles drop-coated on graphene. Intensity profiles crossing three nanoparticles are extracted from the LEEM (b) and eV-TEM (d) image. Typical sizes of the features are 30 nm to 40 nm in both reflection and transmission.

The Au NPs are visible as dark spots in both reflection (2.5 eV electron energy, Fig. 4a) and transmission (4.0 eV, Fig. 4c) mode. Also, the outline of the 2.5 µm hole of the SiN grid underlying the graphene is visible, providing a guide for alignment of the two images. In eV-TEM, the NPs marked by the orange and blue lines are clearly distinguished. The Au NP with the green arrow, however, is not visible against the dark background, highlighting that the 'graphene' here is too thick (>5 layers in this case), which reduces electron transmission dramatically. Clearly, also the SiN grid is non-transparent to these low energy electrons.

In Fig. 4b and 4d, intensity profiles along the lines of corresponding colors are shown (in Fig. 4b for LEEM and in Fig. 4d for eV-TEM, respectively). The sizes of the Au NPs are approx. 40 nm in both imaging modes, leading to the conclusion that we actually see clusters of several Au NPs. Varying the focusing conditions does not lead to significantly smaller features, excluding the possibility that the three-dimensional structure of the NPs acts as an electron lens. We do not observe similar features on samples on which no NPs were deposited.

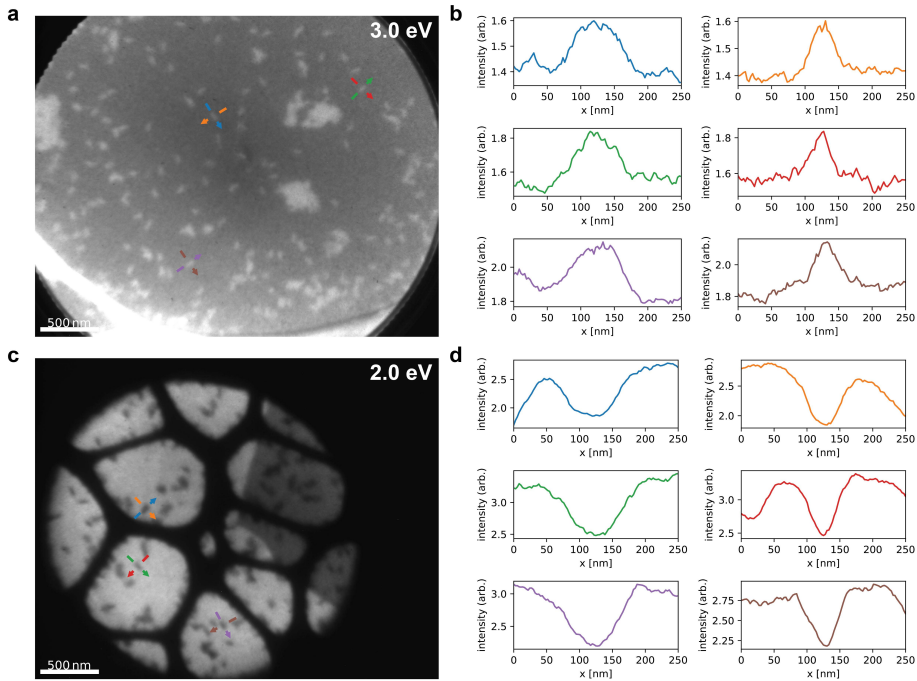

In both modes we analyze line profiles across DNA origami patches, along the short and long axis of the feature. In the line profiles (colors in Fig. 5b correspond to lines in Fig. 5a, and in Fig. 5d to Fig. 5c) the short and long axis of each rectangle can be clearly discerned. The sizes lie in the expected range of 70 nm and 90 nm, respectively.

The eV-TEM micrograph (Fig. 5c) was acquired after illuminating the sample with low energy electrons for one hour at ~3 eV, corresponding to a dose of $\sim 3 \cdot 10^{10}$ electrons per DNA patch. Our findings show that organic samples of tens of nm in size are observable in eV-TEM, without apparent radiation damage over a prolonged time.

*Figure 5: Rectangular DNA origami patches on top of mono- and double-layer graphene oxide suspended over lacey carbon web. LEEM (a) and eV-TEM (c) image of DNA origami on graphene oxide (different areas). The DNA origami is bright in LEEM and dark in eV-TEM. Intensity profiles (b, d) of the rectangular (70 nm by 90 nm) DNA origami show that the short and long axes are distinguishable. In eV-TEM, the DNA origami is partially electron transparent.*

## 5.2 DNA origami on graphene oxide

To demonstrate the applicability of eV-TEM to biological samples, we imaged rectangular DNA origami molecules (Gattaquant Gatta-AFM nanorulers, http://www.gattaquant.com/products/gatta-afm.html). We deposit an aqueous solution with DNA on graphene oxide and let it dry, as graphene oxide is hydrophilic and DNA origami can self-assemble on it. The sample was stored in the microscope load-lock vacuum chamber (pressure in range of $10^{-8}$ mbar) for one week to allow outgassing of residues, as adsorbed water on the hydrophilic surface prevents imaging with low energy electrons (note that heating would disintegrate the DNA).

The DNA origami patches appear as dark spots in the eV-TEM micrograph (Fig. 5c, 2.0 eV). The lacey carbon support is non-transparent to electrons, just like the SiN grids, whereas the graphene oxide covers the hole with two patches of different layer-count. In LEEM (Fig. 5a, 3.0 eV) the small DNA origami patches show up brighter than the graphene oxide.

## 6. Discussion

We have presented transmission micrographs of Au NPs and DNA origami, acquired at electron energies below 10 eV instead of tens to hundreds of keV as in conventional TEM. Thereby, eV-TEM circumvents the problem of electron beam damage to organic and biological samples to a large extent. As eV-TEM is an extension to the standard LEEM setup, we can readily switch between reflection and transmission images of the same area. Sample preparation does not require staining, as shown by depositing objects from solution on graphene and graphene oxide. Furthermore, we have examined the transmission and reflection spectra of few-layer graphene in the 0 to 75 eV energy range. Similar to many other materials ('universal' curve), the transmissivity increases at energies below the MFP minimum at about 30 eV. The characteristic oscillations in the spectra that we attribute to interference of the electron waves is not predicted by the 'universal' MFP curve. Further experiments will show if these oscillations generalize to other multilayer materials as expected for a simple interference phenomenon. It is noteworthy that similar oscillations have also been observed for hBN using LEEM[23]. In addition to their high transmissivity, few-layer van der Waals materials are especially suitable as substrates because they are thin and flat, causing minimal deflection of the transmitted electrons. Below ~28 eV there are no first (or higher) order diffraction beams inside the graphene lattice, so there is only the directly transmitted electron beam, which simplifies analysis of the results. eV-TEM is capable of imaging structures of tens of nm in size, with a resolution better than 10 nm as determined on graphene. The resolution is currently limited by the comparably large energy spread and low spatial coherence of the electron emitter, among other factors. We expect that upgrades to the electron gun will significantly improve the quality of the images, as both electron energy spread and emission area can be dramatically reduced by laser-excited photoemission from a low workfunction transparent cathode.

Thus, eV-TEM provides new opportunities for damage-free imaging in transmission electron microscopy. As it operates in a very different energy range than conventional TEM, we expect new insights into biological matter such as proteins, DNA or cell membranes from eV-TEM. As realization of eV-TEM imaging capabilities requires only very modest modification of the sample holder and electronics, LEEM/PEEM instruments already in operation around the world can adopt this new technology at low cost.

**Acknowledgments**

We thank Johannes Jobst for his scientific and technological input, Kirsten Martens for help with the preparation of the DNA origami, Marcel Hesselberth and Douwe Scholma for their excellent technical support and Ruud van Egmond for building the eV-TEM gun assembly. This work was supported by the Netherlands Organization for Scientific Research (NWO) via the STW-HTSM Grant (No. 12789) and the NWO Vrije Programma (TNW18.071).

**References**

[1]    L.A. Polette, G. Meitzner, M.J. Yacaman, R.R. Chianelli, Maya blue: Application of XAS and HRTEM to materials science in art and archaeology, Microchem. J. 71 (2002) 167–174. https://doi.org/10.1016/S0026-265X(02)00008-5.

[2]    G. Van Tendeloo, High resolution electron microscopy in materials research, J. Mater. Chem. (1998). https://doi.org/10.1039/a708240a.

[3]    R.E. Burge, Mechanisms of contrast and image formation of biological specimens in the transmission electron microscope, J. Microsc. (1973). https://doi.org/10.1111/j.1365-2818.1973.tb03832.x.

[4]    P.W. Hawkes, J.C.H. Spence, Springer Handbook of Microscopy, 2019.

[5]    A. V. Crewe, J. Wall, J. Lanomore, Visibility of single atoms, Science (80-. ). (1970). https://doi.org/10.1126/science.168.3937.1338.

[6]    J.C. Meyer, C.O. Girit, M.F. Crommie, A. Zettl, Imaging and dynamics of light atoms and molecules on graphene, Nature. 454 (2008) 319–322. https://doi.org/10.1038/nature07094.

[7]    R. Henderson, Realizing the potential of electron cryo-microscopy, Q. Rev. Biophys. 37 (2004) 3–13. https://doi.org/10.1017/S0033583504003920.

[8]    M. Linck, P. Hartel, S. Uhlemann, F. Kahl, H. Müller, J. Zach, M. Haider, M. Niestadt, M. Bischoff, J. Biskupek, Z. Lee, T. Lehnert, F. Börrnert, H. Rose, U. Kaiser, Chromatic Aberration Correction for Atomic Resolution TEM Imaging from 20 to 80 kV, Phys. Rev. Lett. 117 (2016) 1–5. https://doi.org/10.1103/PhysRevLett.117.076101.

[9]    U. Kaiser, J. Biskupek, J.C. Meyer, J. Leschner, L. Lechner, H. Rose, M. Stöger-Pollach, A.N. Khlobystov, P. Hartel, H. Müller, M. Haider, S. Eyhusen, G. Benner, Transmission electron microscopy at 20kV for imaging and spectroscopy, Ultramicroscopy. 111 (2011) 1239–1246. https://doi.org/10.1016/j.ultramic.2011.03.012.

[10]    M.P. Seah, W.A. Dench, Quantitative electron spectroscopy of surfaces: A

standard data base for electron inelastic mean free paths in solids, Surf. Interface Anal. (1979). https://doi.org/10.1002/sia.740010103.

[11] I. Müllerová, M. Hovorka, R. Hanzlíková, L. Frank, Very Low Energy Scanning Electron Microscopy of Free-Standing Ultrathin Films, Mater. Trans. 51 (2010) 265–270. https://doi.org/10.2320/matertrans.MC2009 15.

[12] J.N. Longchamp, T. Latychevskaia, C. Escher, H.W. Fink, Low-energy electron holographic imaging of individual tobacco mosaic virions, Appl. Phys. Lett. 107 (2015). https://doi.org/10.1063/1.4931607.

[13] G.B. Stevens, M. Krüger, T. Latychevskaia, P. Lindner, A. Plückthun, H.W. Fink, Individual filamentous phage imaged by electron holography, Eur. Biophys. J. 40 (2011) 1197–1201. https://doi.org/10.1007/s00249-011-0743-y.

[14] D. Geelen, A. Thete, O. Schaff, A. Kaiser, S.J. van der Molen, R. Tromp, eV-TEM: Transmission electron microscopy in a low energy cathode lens instrument, Ultramicroscopy. 159 (2015) 482–487. https://doi.org/10.1016/j.ultramic.2015.06.0 14.

[15] D. Geelen, eV-TEM: Transmission Electron Microscopy with Few-eV Electrons, 2018. http://hdl.handle.net/1887/63484

[16] D. Geelen, J. Jobst, E.E. Krasovskii, S.J. van der Molen, R.M. Tromp, Nonuniversal Transverse Electron Mean Free Path through Few-layer Graphene, Phys. Rev. Lett. 123 (2019) 86802. https://doi.org/10.1103/physrevlett.123.086 802.

[17] S.M. Schramm, J. Kautz, A. Berghaus, O. Schaff, R.M. Tromp, S.J. Van Der Molen, Low-energy electron microscopy and spectroscopy with ESCHER: Status and prospects, in: IBM J. Res. Dev., 2011. https://doi.org/10.1147/JRD.2011.2150691.

[18] S.M. Schramm, A.B. Pang, M.S. Altman, R.M. Tromp, A Contrast Transfer Function approach for image calculations in standard and aberration-corrected LEEM and PEEM, Ultramicroscopy. 115 (2012) 88–108. https://doi.org/10.1016/j.ultramic.2011.11.0 05.

[19] R.M. Tromp, J.B. Hannon, A.W. Ellis, W. Wan, A. Berghaus, O. Schaff, A new aberration-corrected, energy-filtered LEEM/PEEM instrument. I. Principles and design, Ultramicroscopy. 110 (2010) 852–861. https://doi.org/10.1016/j.ultramic.2010.03.0 05.

[20] W. Regan, N. Alem, B. Alemán, B. Geng, Ç. Girit, L. Maserati, F. Wang, M. Crommie, A. Zettl, A direct transfer of layer-area graphene, Appl. Phys. Lett. 96 (2010) 2008–2011. https://doi.org/10.1063/1.3337091.

[21] K.S. Novoselov, A.K. Geim, S. V. Morozov, D. Jiang, Y. Zhang, S. V. Dubonos, I.V. Grigorieva, A.A. Firsov, Electric Field Effect in Atomically Thin Carbon Films, Science (80-. ). 306 (2004) 666–669. https://doi.org/10.1126/science.1102896.

[22] J. Jobst, J. Kautz, D. Geelen, R.M. Tromp, S.J. Van Der Molen, Nanoscale measurements of unoccupied band dispersion in few-layer graphene, Nat. Commun. 6 (2015) 1–6. https://doi.org/10.1038/ncomms9926.

[23] J. Jobst, A.J.H. Van Der Torren, E.E. Krasovskii, J. Balgley, C.R. Dean, R.M. Tromp, S.J. Van Der Molen, Quantifying electronic band interactions in van der Waals materials using angle-resolved reflected-electron spectroscopy, Nat. Commun. 7 (2016) 1–6. https://doi.org/10.1038/ncomms13621.

[24] H. Anders, Thin Films in Optics, Focal P., 1967.

[25] R.M. Feenstra, M. Widom, Low-energy electron reflectivity from graphene: First-principles computations and approximate models, Ultramicroscopy. 130 (2013) 101–108. https://doi.org/10.1016/j.ultramic.2013.02.0 11.

[26] E.E. Krasovskii, Augmented-plane-wave approach to scattering of Bloch electrons

by an interface, Phys. Rev. B - Condens. Matter Mater. Phys. 70 (2004) 1–11. https://doi.org/10.1103/PhysRevB.70.2453 22.

[27] V.U. Nazarov, E.E. Krasovskii, V.M. Silkin, Scattering resonances in two-dimensional crystals with application to graphene, Phys. Rev. B - Condens. Matter Mater. Phys. 87 (2013) 1–5. https://doi.org/10.1103/PhysRevB.87.0414 05.

[28] Y.J. Yu, Y. Zhao, S. Ryu, L.E. Brus, K.S. Kim, P. Kim, Tuning the graphene work function by electric field effect, Nano Lett. 9 (2009) 3430–3434. https://doi.org/10.1021/nl901572a.

[29] R.R. Nair, P. Blake, J.R. Blake, R. Zan, S. Anissimova, U. Bangert, A.P. Golovanov, S. V. Morozov, A.K. Geim, K.S. Novoselov, T. Latychevskaia, Graphene as a transparent conductive support for studying biological molecules by transmission electron microscopy, Appl. Phys. Lett. (2010). https://doi.org/10.1063/1.3492845.